\documentclass[amsmath,amssymb,
 reprint]{revtex4}

\usepackage{graphicx}% Include figure files
\usepackage{dcolumn}% Align table columns on decimal point
\usepackage{bm}% bold math
\usepackage[utf8]{inputenc}
\usepackage[T1]{fontenc}
\usepackage{mathptmx}
\usepackage{hyperref}
\usepackage{mathrsfs}
\usepackage{amsmath}
\usepackage{amsfonts}
\usepackage{amssymb}
\usepackage{graphicx}
\usepackage[dvipsnames]{xcolor}
\usepackage{subcaption}
\usepackage[normalem]{ulem}

\usepackage{xcolor}

\newcommand{\VERSIONS}[1]{}

\newcommand{\comment}[1]{}

\usepackage{amsmath}

\renewcommand{\arraystretch}{1.5}

\begin{document}

%\preprint{AIP/123-QED}

\title{
Subdiffusive-Brownian crossover in membrane proteins: a Generalized Langevin Equation-based approach
}

\author{Loris Di Cairano}
 \email{l.di.cairano@fz-juelich.de}
 \affiliation{Department of Physics, Faculty of Mathematics, Computer Science and Natural Sciences, Aachen University, 52062 Aachen, Germany}
\affiliation{%
Computational Biomedicine, Institute of Neuroscience and Medicine INM-9 and Institute for Advanced Simulations IAS-5, Forschungszentrum Jülich, 52428 Jülich, Germany%\\This line break forced% with \\
}%
% \affiliation{ 
% Applied and Computational Mathematics, Department of Mathematics, RWTH Aachen University, Aachen, Germany%\\This line break forced with \textbackslash\textbackslash
\author{Benjamin Stamm}%
 \email{best@acom.rwth-aachen.de}
\affiliation{ 
Applied and Computational Mathematics, Department of Mathematics, RWTH Aachen University, Aachen, Germany%\\This line break forced with \textbackslash\textbackslash
}%

\author{Vania Calandrini}
%  \homepage{http://www.Second.institution.edu/~Charlie.Author.}
 \email{v.calandrini@fz-juelich.de}
\affiliation{%
Computational Biomedicine, Institute of Neuroscience and Medicine INM-9 and Institute for Advanced Simulations IAS-5, Forschungszentrum Jülich, 52428 Jülich, Germany%\\This line break forced% with \\
}%

\date{\today}% It is always \today, today,

\begin{abstract}
In this paper, we propose a Generalized Langevin Equation (GLE)-based model to describe the lateral diffusion of a protein in a lipid bilayer. The memory kernel is represented in terms of a viscous (instantaneous) and an elastic (non instantaneous) component modeled respectively through  a  Dirac  delta  function and  a three-parameter Mittag-Leffler type function.  By imposing a specific relationship between the parameters of the three-parameters Mittag-Leffler  function,  the  different  dynamical  regimes, namely ballistic,  subdiffusive  and Brownian, as well as the crossover from one regime to another, are  retrieved. Within this approach, the transition time from the ballistic to the subdiffusive regime and the distribution of relaxation times underlying the transition from the subdiffusive to the Brownian regime are given. The reliability of the model is tested by comparing the Mean Squared Displacement (MSD) derived in the framework of this model and the MSD of a protein diffusing in a membrane calculated through molecular dynamics (MD) simulations.

\end{abstract}

% \tableofcontents
\maketitle

\section*{Significance}

This study reports a Generalized Langevin Equation (GLE) model, based on a three-parameter Mittag-Leffler memory kernel, to describe a protein laterally diffusing in a membrane. The model captures the different  dynamical regimes, namely ballistic,  subdiffusive  and Brownian, as well as the crossover between them.  The distribution of relaxation times underlying the transition from the subdiffusive to the Brownian regime is given. 

\section*{Introduction}
Lateral diffusion in membrane is key for cellular information
processing \cite{diffusion_biology}.
Cell membrane fluidity determines lipid and protein mixing, thus regulating diffusion-limited biochemical interaction rates responsible for signal transduction from the extracellular to the intracellular environment. 
\\
In the last years, it has been studied how biological structures and features of living cells, such as membrane composition, concentration of proteins in membrane, compartmentalization and crowding, affect the diffusion of proteins or lipids in membrane \cite{concentration_protein,compartment_membrane}. A plethora of experimental studies in cellular membranes \cite{fujiwara2002phospholipids,murase2004ultrafine,weigel2011ergodic,gielen2005diffusion,gupta2018dynamics}, and in vitro lipid bilayers \cite{horton2010development,ramadurai2009lateral,deverall2005membrane}, as well as computer simulations of minimalistic model membranes in crowding conditions \cite{jeon2016protein,sugar2005lateral,skaug2011correlating,kammerer2008cluster,saxton1994anomalous} have shown a deviation from the simple Brownian diffusion, where the random displacements are described by a Gaussian probability distribution and the mean squared displacement (MSD) increases linearly in time when time-lags much longer than the typical collision time are considered ($MSD \propto D t$, where $D$ is the diffusion constant, $D=K_{B}T/\mu$, with $K_{B}$, $T$ and $\mu$ being respectively the Boltzmann constant, the temperature, and a constant accounting for the geometry of the particle and the cinematic viscosity of the environment). 
 One of the most familiar phenomena is indeed a sub-linear, power-law increase of the $MSD \propto D_\alpha t^\alpha$, with $0 < \alpha < 1$ and $D_\alpha$ representing a diffusion constant with the physical dimensions of $[L^2/t^\alpha]$. 
\\
Such \emph{subdiffusive} dynamics, also referred to as anomalous dynamics, is commonly attributed to the densely packed and heterogeneous structures resulting from the crowding of the biological membranes. Notably, atomistic and coarse-grained simulations, as well as this study, have shown subdiffusive behavior also for lipids and proteins in simple lipid bilayers, in the limit of infinite protein dilution \cite{anomalous_diff_cholesterol,crowded_membrane,kneller2014communication}. Moreover, as observed in many viscoelastic systems, subdiffusivity is a transient dynamical behavior \cite{hofling_anomalous}. After the short-time ballistic regime, where the tagged particle freely diffuses ($MSD \propto (K_{B}T/M) t^2$, with $M$ being the the particle's mass), the interactions with the medium and its specific fluidic/mechanical properties can lead to non-trivial persistent correlations responsible for the subdiffusive behavior \cite{saxton2007biological}. Yet, for time-lags exceeding some characteristic time, the standard Brownian dynamics is recovered. This Brownian regime is well described by the celebrated Saffman and Delbrück model \cite{saffman1975brownian,saffman1976brownian} and its extensions \cite{hughes1981translational,block2018brownian,Naji2007,Venable2017,Fournier2019}. The crossover from the subdiffusive to the standard Brownian dynamics can take place over a quite large time window and the transition onset strongly depends on packing and crowding, ranging from tenth to hundreds of nanoseconds for lipids in protein-free membranes or proteins in membranes at infinite protein dilution, up to arbitrarily long time scales for crowded real systems \cite{crowded_membrane,hofling_anomalous,fractional_msd}. As a consequence, diffusive properties can no longer be characterized by a single diffusion constant nor by a single exponent $\alpha$.
\\
Several theoretical frameworks have been introduced to reproduce the subdiffusive behavior and describe the physical mechanism behind it (see for instance Ref. \citep{hofling_anomalous} and references therein for a complete overview on their theoretical foundations and applications to biological systems).
Among them, we recall the so-called Gaussian models, such as the \emph{Fractional Brownian motion} and the \emph{Generalized Langevin Equation} (GLE) \cite{hofling_anomalous}, where the statistics of the noise in the relevant stochastic equations is still assumed to be Gaussian like in the Langevin approach for normal Brownian diffusion, but, differently from this, the noise displays persistent power-law correlations (power-law Gaussian noise) leading to a sub-linear increase of the MSD. 
In the Generalized Langevin Equation, due to the fluctuation-dissipation theorem, this amounts to generalize the Stokes drag term to a convolution of the velocity of the particle with a power-law memory kernel representing a generalized time-dependent friction \cite{zwanzig}.
Another important class is the one of the \emph{Continuous-time random walks} \cite{hofling_anomalous}, where particles undergo a series of displacements in an homogeneous medium according with a waiting-time distribution. Anomalous transport is connected with a power-law tail in the waiting-time distribution, such that even the mean-waiting time is infinite. The central-limit theorem does not apply in this case, since longer and longer waiting times are sampled. 
Finally, another important category is represented by the \emph{Lorentz models} \cite{hofling_anomalous} that consider spatially disordered environments where the
particle explores fractal-like structures, leading to anomalous dynamics.\\
Recently, in the framework of the generalized Langevin equation model, hard exponential and soft power-law truncation (tempering) of power-law memory kernels \cite{Metzler}, as well as exponentially truncated three parameter Mittag Leffler-based memory kernels \cite{liemert2017generalized} have been used in order to quantitatively reproduce the transition from subdiffusive to normal diffusion. Both exponential and soft power-law truncation imply the introduction of a characteristic cross-over time, related to a maximum-correlation time in the driving noise. Applications of these models to the analysis of the MSD of lipids in simple lipid bilayers have shown that the cross-over time is related to the time scale of mutual exchange between lipids \cite{Metzler,jeon2012anomalous}.
Compared to simply combining an anomalous and a normal diffusion law for the MSD, such phenomenological models naturally yield the emergence of subdiffusive to normal crossover dynamics into a unique model, where the type of truncation governs the crossover shape. 
\\
% \cbs{Loris, insert here the references about ML-functions of visco-elastic phenomenons in a different context}
%The open challenge is to understand what are the relevant degrees of freedom and environment's properties responsible for such phenomenological kernels.
%
To address this problem in the case of laterally diffusing membrane proteins, here we propose a GLE-based model where the memory kernel is borrowed from a viscoelastic representation of the lipid membrane \cite{viscoelastic_membrane,viscoelastic_relaxation,viscoelastic_breakdown,viscoelastic_bending,hofling_anomalous,rheology_lipid,LevineMacKintosh2002}. Specifically, we represent the kernel in terms of a viscous (instantaneous) and an elastic (non instantaneous) term modeled respectively through a Dirac delta function and the solution of the Prabhakar fractional derivative \cite{d2013fractional,caputo_derivative}, i.e. a three-parameter Mittag Leffler-based function \cite{prabhakar1971singular,convergence_mittag_leffler}.
We note that the three-parameter Mittag-Leffler function has also been used in different contexts to model viscoelastic effects \cite{PhysRevLett.102.058101,goychuk2012viscoelastic,de2011models}.
We show that imposing a specific relationship between the parameters of the three-parameters Mittag Leffler function, naturally yields the emergence of the different dynamical regimes of the protein (ballistic, subdiffusive and Brownian) and the crossover between them, with a well defined finite diffusion coefficient, without introducing hard exponential truncation function (used in the mathematical study of Ref. \cite{liemert2017generalized}). The distribution of relaxation times underlying the transition from the subdiffusive to the Brownian regime is derived.
The reliability of the proposed GLE-model is tested versus the MSD data of a protein (the muscarinic M2 receptor) embedded in a mixed membrane, produced by MD simulations. \\
From a continuum perspective, the reliability of the proposed kernel in describing the transition among the different dynamical regimes, suggests this function as a possible candidate to describe the time-dependent membrane response within the constitutive equation (as done in Ref. \cite{free_confined} for a simpler model).
This topic will be address in a forthcoming paper  together with a study on the dependence of the Mittag-Leffler parameters on the membrane composition.

\section*{Methods}

% {\color{green}I propose to simplify the subsections.}
%\subsection{Theoretical Model}
%\label{sec:theory}

%\subsubsection{Generalized Langevin Equation for Protein Diffusion}
%\label{ssec:GLE_for_protein}
\subsection*{Generalized Langevin Equation for Protein Diffusion}
\label{ssec:GLE_for_protein}
Our aim is to model the transition from ballistic, subdiffusive, to Brownian  motion of a protein  diffusing laterally in a lipid bilayer in terms of  a Generalized Langevin Equation of the form
\begin{equation}
\begin{split} 
    \label{eq:GLE_start}
    M \frac{d\bm{U}_{c}}{dt}(t)&=-\int_{0}^{t} \zeta(t-u) \bm{U}_{c}(u)\,du+\bm{\Xi}(t),
\end{split}
\end{equation}
where $\bm{U}_{c}(t)\in \mathbb R^2$ denotes the protein's center of mass velocity in the xy-plane, $\zeta$ the kernel function and $\bm{\Xi}$ is a two-dimensional noise term.
Here, we model the friction term by
\begin{equation}\label{def:viscosity_func_viscoelastic}
    \zeta(t):=\xi_{s}\delta(t)+\xi_{p}\frac{t^{\nu-1}}{\tau^{\nu}}E^{\delta}_{\lambda,\nu}\left[-\left(\frac{t}{\tau}\right)^{\lambda}\right],
\end{equation}
where $E^{\delta}_{\lambda,\nu}$ denotes the 3-parameter Mittag-Leffler function, also called Prabhakar function \cite{prabhakar1971singular,convergence_mittag_leffler}.
For physical reason, in order to have a monotonous kernel function, $\zeta$, which ensures a monotonous energy decay in isolated systems and a non-negative spectral distribution, the parameters of the Mittag-Leffler function have to verify the relation \cite{mainardi2015complete} $ 0<\lambda\leq 1$, and $0<\lambda\delta\leq\nu\leq 1$.
We refer to Appendix~\ref{appex:ML} for further details on the 3-parameter Mittag-Leffler function including an illustration and asymptotic behavior.
The parameters $\xi_{s}$ and $\xi_{p}$ have the dimension of a friction and account for the contributions coming from the instantaneous (viscous) and retarded (elastic) response, respectively.

The $\{\Xi^{i}(t)\}_{i=x,y}$ are the components of a Gaussian distributed colored thermal noise that verifies  the fluctuation-dissipation theorem:
\begin{equation}
    \langle \Xi^{i}(t)\;\Xi^{j}(u)\rangle=2K_{B}T\,\zeta(t-u)\delta_{ij}, \qquad i=x,y.
\end{equation}

%\subsection{Theoretical Results}
%\subsubsection{Solution of the Model and statistical observables}\label{ssec:solution_ECM}
\subsection*{Solution of the Model and Statistical Observables}\label{ssec:solution_ECM}

In order to solve the GLE model \eqref{eq:GLE_start}, we adopt standard methods for Stochastic differential equations as used in Ref.
\cite{convergence_mittag_leffler}. 
Applying the Laplace transform to Eq. \eqref{eq:GLE_start} and denoting the Laplace transformed functions with the overscript $\;\hat{}\;$, it reduces to:
\begin{equation}\label{eqn:velocity_algebraic_GLE}
    \hat{\bm{U}}_{c}(s)=M\,\bm{U}_{c}(0)\,\hat{g}(s)+\hat{\bm{\Xi}}(s)\,\hat{g}(s),
\end{equation}
for the velocity vector. 
Equivalently for the position vector, ($\bm{U}_{c}=\dot{\bm{X}}_{c}$) it is
\begin{equation}\label{eqn:position_algebraic_GLE}
    \hat{\bm{X}}_{c}(s)=\frac{\bm{X}_{c}(0)}{s}+M\,\bm{U}_{c}(0)\hat{H}(s)+\hat{\bm{\Xi}}(s)\,\hat{H}(s).
\end{equation}
Here, we have defined the so-called $\hat{g}(s)$ and $\hat{H}(s)$-\emph{relaxation functions}, respectively, given by:
\begin{equation}\label{g_RF}
    \hat{g}(s):=\frac{1}{Ms+\hat{\zeta}(s)},
\end{equation}
\begin{equation}\label{H_RF}
    \hat{H}(s):=\frac{\hat{g}(s)}{s}=\frac{1}{Ms^{2}+s \, \hat{\zeta}(s)},
\end{equation}
where the Laplace transform of $\zeta$ writes \cite{convergence_mittag_leffler}:
\begin{equation}
\label{zetaS}
\begin{split}
        \hat{\zeta}(s):=\xi_{s}+\xi_{p}\frac{(\tau s)^{\delta\lambda-\nu}}{(1+(\tau s)^{\lambda})^{\delta}}.
\end{split}
\end{equation}
By applying the inverse Laplace transform to Eqs. \eqref{eqn:velocity_algebraic_GLE} and \eqref{eqn:position_algebraic_GLE}, one obtains the formal solution for, respectively, the velocity and position vectors by
\begin{equation}\label{solution_gle_velocity}
\begin{split}
        \bm{U}_{c}(t)&=\langle\bm{U}_{c}(t)\rangle+\int_{0}^{t}g(t-u)\,\bm{\Xi}(u)\,du,\\
        \langle\bm{U}_{c}(t)\rangle&=M\,\bm{U}_{c}(0)\,g(t),
\end{split}
\end{equation}
and
\begin{equation}
\begin{split}\label{solution_gle_position}
        \bm{X}_{c}(t)&=\langle\bm{X}_{c}(t)\rangle+\int_{0}^{t}H(t-u)\bm{\Xi}(u)\,du,\\
        \langle\bm{X}_{c}(t)\rangle&=\bm{X}_{c}(0)+M\,\bm{U}_{c}(0)\,H(t).
\end{split}
\end{equation}
In order to obtain an analytical expression for the MSD, we finally introduce the $I$-\emph{relaxation function}, defined by:
\begin{equation}
\label{I_RF}
    \hat{I}(s)=\frac{\hat{g}(s)}{s^{2}}=\frac{1}{Ms^{3}+s^{2}\hat{\zeta}(s)}.
\end{equation}
By analytically computing the inverse Laplace transform of the relaxation functions $g,\,H$ and $I$, the stochastic process is also analytically solved. In fact, one can show that the statistical observables such as MSD, VACF and time-dependent diffusion coefficient, $D(t)$, are related to the relaxation functions through the following relations \cite{vinales2007anomalous,correlation_method_vinales,computation_correlation_function,correlation_method_sandev,convergence_mittag_leffler}:
\begin{equation}
    \begin{split}\label{def:stat_obs}
        C_{\textbf{v}}(t)&=2 K_{B}T\, g(t),\\
        D(t)&=K_{B}T\, H(t),\\
        \langle \Delta\bm{X}^{2}(t)\rangle &=4\,K_{B}T\, I(t).
    \end{split}
\end{equation}

\subsection*{Asymptotic Behaviors of Statistical Observables}\label{ssec:asymptotic_behaviors}

%\subsubsection{Short and long time behaviors of correlation functions}

The short time behavior of the kernel function $\zeta(t)$ defined by \eqref{def:viscosity_func_viscoelastic} can be obtained by taking the first term in the series representation of the Mittag-Leffler function (Eq.~\eqref{def_ML_3par} in Appendix~\ref{appex:ML}) which is
\begin{equation}
    \label{ML_short_time}
    E^{\delta}_{\lambda,\nu}(-z)\approx \frac{1}{\Gamma(\nu)}, \qquad (|z|\ll1),
\end{equation}
and we obtain the short time kernel function $\zeta_{S}$ by
\begin{equation}
    \zeta_{S}(t)=\xi_{s}\delta(t)+\frac{\xi_{p}}{\tau^{\nu}}\frac{t^{\nu-1}}{\Gamma(\nu)}, \qquad (t\ll\tau)
\end{equation}
whose Laplace transform reads
\begin{equation}
    \hat{\zeta}_{S}(s)= \xi_{s}+\frac{\xi_{p}}{\tau^{\nu}} s^{-\nu}.
\end{equation}
By substituting the asymptotic kernel function in the relaxation functions, we get the short time asymptotic relaxation functions:
\begin{equation}
    \label{eq:short_time_RF}
    \begin{split}
        \hat{g}_{S}(s)&:=\frac{1}{M s+\xi_{s}+\frac{\xi_{p}}{\tau^{\nu}}s^{-\nu}},\\
        \hat{H}_{S}(s)&:=\frac{1}{M s^{2}+\xi_{s}s+\frac{\xi_{p}}{\tau^{\nu}}s^{-\nu+1}},\\
        \hat{I}_{S}(s)&:=\frac{1}{M s^{3}+\xi_{s}s^{2}+\frac{\xi_{p}}{\tau^{\nu}}s^{-\nu+2}}.
    \end{split}
\end{equation}
By adopting the methods introduced in Refs. \cite{ift_vacf,ift_vacf_hh}, we can apply the inverse Laplace transform so to obtain the analytic behavior of the statistical observables at short time in accordance with definitions \eqref{def:stat_obs}, namely:
\begin{equation}
    \begin{split}\label{}
      C^{S}_{\bm{v}}(t)&\approx\frac{2K_{B}T}{M}\Bigg[1-\omega_{s}t\Bigg], 
      \\
      D^{S}(t)
      &\approx
      \frac{\phantom{2}K_{B}T}{M} \Bigg[1-\frac{\omega_{s} t}{2}\Bigg] \, t,
      \\
      \langle\Delta\bm{X}^{2}\rangle^{S}(t)
      &\approx
      \frac{2K_{B}T}{M}\Bigg[1-\frac{ \omega_{s}t }{3}\Bigg] \, t^{2},
    \end{split}
\end{equation}
where 
\begin{equation}
    \omega_{s}:=\frac{\xi_{s}}{M},
\end{equation}
is a characteristic frequency defining the time scale of the transition from the ballistic to subdiffusive regime.

It is also useful to introduce the parameter
\begin{equation}
\omega_{p}:=\frac{\xi_{p}}{M},
\end{equation}
corresponding to $\xi_p$.
\\
The long time behavior can be obtained by imposing the following inequality in the relaxation functions
\[
 \frac{M}{\xi_{s}} s\ll 1,
\]
that applied to Eqs. \eqref{g_RF}, \eqref{H_RF} and \eqref{I_RF} respectively yields the asymptotic relaxation functions
\begin{equation}\label{long_time_relax_functs}
    \begin{split}
        \hat{g}_{B}(s)&:=\frac{1}{\xi_{s}}\left(\frac{1}{1+\frac{\phi}{\tau^{\nu}}\frac{s^{\delta\lambda-\nu}}{(\tau^{-\lambda}+s^{\lambda})^{\delta}}}\right),\\
        \hat{H}_{B}(s)&:=\frac{1}{\xi_{s}}\left(\frac{1}{s+\frac{\phi}{\tau^{\nu}}\frac{s^{\delta\lambda-\nu+1}}{(\tau^{-\lambda}+s^{\lambda})^{\delta}}}\right),\\
        \hat{I}_{B}(s)&:=\frac{1}{\xi_{s}}\left(\frac{1}{s^{2}+\frac{\phi}{\tau^{\nu}}\frac{s^{\delta\lambda-\nu+2}}{(\tau^{-\lambda}+s^{\lambda})^{\delta}}}\right),
    \end{split}
\end{equation}
where we introduced the dimensionless parameter $\phi=\xi_{p}/\xi_{s}$.\\
As for the short time behavior, we adopt the same methods introduced in Refs. \cite{ift_vacf,ift_vacf_hh} so that the inverse Laplace transforms of the above relaxation functions lead to the analytical expression for the statistical observables
\begin{equation}
\label{very_long_time_RF}
\begin{split}
     C^{B}_{\bm{v}}(t)
     &\approx 
     \frac{\tau^{-\delta\lambda+\nu}}{ \xi_{p}}%\phi \, \xi_{s}}
     \frac{t^{\delta\lambda-\nu-1}}{\Gamma(\delta\lambda-\nu)}
     -\frac{\tau^{\lambda}}{ \xi_{p}}%\phi\xi_{s}}
     \frac{t^{-\lambda-1}}{\Gamma(-\lambda+(\delta\lambda-\nu))},
     \\
    D^{B}(t)
    &\approx \frac{\tau^{-\delta\lambda+\nu}}{ \xi_{p}}%\phi \, \xi_{s}}
    \frac{t^{\delta\lambda-\nu}}{\Gamma((\delta\lambda-\nu)+1)},
    \\
    \langle\Delta\bm{X}^{2}\rangle^{B}(t)
    &\approx \frac{\tau^{-\delta\lambda+\nu}}{ \xi_{p}}%\phi \, \xi_{s}}
    \frac{t^{\delta\lambda-\nu+1}}{\Gamma((\delta\lambda-\nu)+2)}.
\end{split}
\end{equation}
\\
In order to recover the Brownian regime in the long time limit, $D(t)$ has to approach a constant $D_\infty > 0$.
From the second Eq. in \eqref{very_long_time_RF}, one notices that this requirement is satisfied if and only if:
\begin{equation}\label{Eq:constraint_exponent}
    \delta\lambda-\nu=0.
\end{equation}
The exact mathematical expression for $D_\infty$ can be obtained considering the non approximated expression for $\hat{H}(s)$  defined by \eqref{H_RF} that, under the constraint \eqref{Eq:constraint_exponent}, reads  
\[
    \hat{D}(s)=\frac{K_{B}T}{s}\left(\frac{1}{M\,s +\xi_{s}+\frac{\xi_{p}}{(1+(\tau\,s)^{\lambda})^{\delta}}}\right).
\]
By taking the limit $s\to 0$, 
the leading term is
\[
    \hat{D}(s)\approx \frac{K_{B}T}{s}\frac{1}{\xi_{s}+\xi_{p}}.
\]
The inverse Laplace transform gives then the exact expression for the asymptotic diffusion coefficient
\begin{equation}
\label{eq:exactDinfty}
    D_\infty=\frac{K_{B}T}{\xi_{s}+\xi_{p}}.
\end{equation}
By applying the constraint \eqref{Eq:constraint_exponent} on expressions \eqref{very_long_time_RF}, we finally get the asymptotic behavior of  the VACF:
\begin{equation}
        C_v^B(t)\approx -\frac{\tau^{\lambda}}{\xi_{p}}\frac{t^{-\lambda-1}}{\Gamma(-\lambda)},
\end{equation}
where we used $\lambda\notin \mathbb{Z}$ and in the same manner for the MSD:
\begin{equation}
    \langle\Delta \bm{X}(t)^{2}\rangle\approx\frac{4 K_{B}T}{\xi_{p}}t,
\end{equation}
so that the linear Brownian regime in the long time is recovered. Notice that in this limit, the friction is dominated by the part $\xi_p$.
Finally, we summarize the physical role of each parameter in Table~\ref{tab:summary_phys_role}.

%%%%%%%%%%%%% Table %%%%%%%%%%%%% 
\begin{table*}[t]
\small
\centering
\begin{tabular}{|l|l|}
\hline
\textbf{Parameters} & \textbf{Physical Role} \\
\hline
$\xi_{s}$ &  friction component coming from the instantaneous response\\
\hline
$\xi_{s}/M$ & characteristic frequency setting the time scale of the transition from ballistic to subdiffusive regime\\
\hline
$\xi_{p}$ & friction component coming from the retarded response. It is the leading term in the asymptotic diffusion coefficient
\\
\hline
$\xi_{p}+\xi_{s}$ & total macroscopic friction felt by the protein\\
\hline
$\tau$ & time scale of the retarded (elastic) response of the lipid membrane\\
\hline
$\delta,\;\lambda,\;\nu $ & $\delta\lambda=\nu$ to  asymptotically get the Brownian regime. They shape the transition from subdiffusive to Brownian regime\\
\hline
\end{tabular}
\caption{Summary of the physical role of the parameters.}
\label{tab:summary_phys_role}
\end{table*}
%%%%%%%%%%%%% End Table %%%%%%%%%%%%% 

%\subsection{Simulation}
%\label{sec:simulation}
\section*{Results and Discussion}
\label{sec:simulation}

As a proof of concept, we test the model against in-silico MSD data of the center of mass of a protein laterally diffusing in a fully hydrated membrane.  The in-silico MSD data are extracted from quasi-atomistic Martini \cite{qi2015charmm,marrink2007martini,marrink2004coarse} MD simulations. The model is thus fitted to the numerical data in order to estimate the best values of the model parameters.

%%%%%%%%%%%%% Figure %%%%%%%%%%%%% 
\begin{figure}[htp!]
  \centering
    \includegraphics[width=0.45\linewidth]{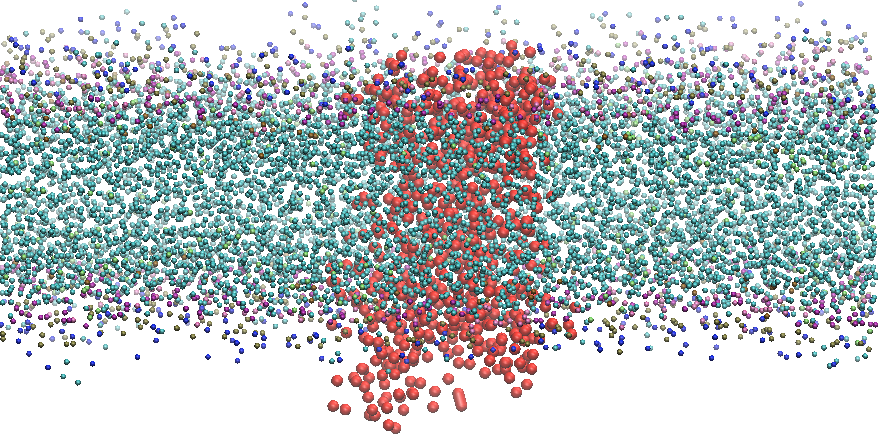}
   \caption{Snapshot of the simulated system. Water and ions are not illustrated for the sake of clarity.}
    \label{fig:3uon_without_loop}
\end{figure}
%%%%%%%%%%%%% End Figure %%%%%%%%%%%%% 

%\subsubsection{System and MD simulations setup}
%\label{ssec:methods}
\subsection*{System and MD Simulations Setup}
\label{ssec:methods}

We used as model system the M2 muscarinic acetylcholine receptor (M2 receptor), an inhibitory class A G-protein-coupled receptor expressed both in the central and parasympathetic nervous systems \cite{haga2012structure}. The inactive form of M2 \cite{haga2012structure} (pdb 3UON) was embedded in a mixed lipid bilayer containing some of the most abundant species in neuronal cell membranes \cite{chan2012comparative}. The antagonist and the fusioned T4 lysozyme, which in the original pdb structure replaces the third intracellular loop of the receptor,  were removed from the system. About 14000 water molecules were added as well as Na$^+$ and Cl$^-$ ions in order to neutralize the system and reach a physiological concentration of 0.15 M. A snapshot of the simulated system is shown in Fig. \ref{fig:3uon_without_loop}. The membrane composition and the characteristic size of the system as well as the simulation details are reported in Appendix~\ref{appex:numerics}. 

\subsection*{Model {\it vs} MD Simulations}
%\section{Results and Discussion}
%\subsection{Numerical Results}

% Old name: Model versus numerical data
\label{ssec:comparison}

\begin{figure}[htp!]
\centering
        \includegraphics[width=0.6\linewidth]{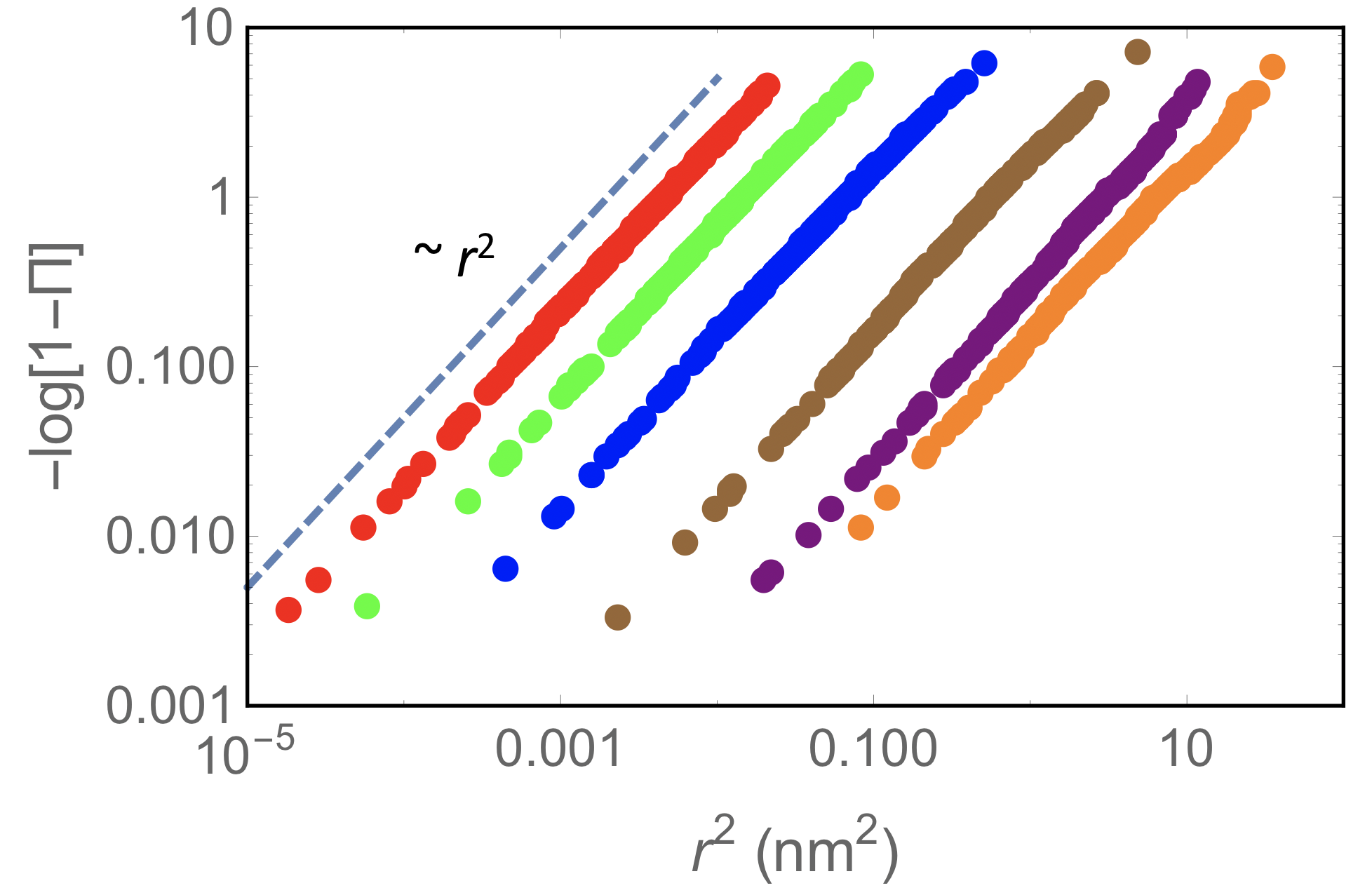}
        \caption{Cumulative distribution $\Pi(r^{2},\Delta)$ with $\Delta=1,10,100,1000,5000,10000$ ns (from left to right). The dashed line corresponds to a theoretical curve proportional to~$r^2$.}
        \label{fig:cumulative_distr}
\end{figure}

The protein dynamics in lipid membrane has been observed to be non Gaussian in case several cases such as, for instance, compartmentalization and crowding \cite{jeon2016protein}. In our case we work at infinite protein dilution, but the membrane is mixed and contains cholesterol, which could cause local inhomogeneity. As a first step we thus verified 
if sizable deviation from a Gaussian process are observed. By adopting the same approach as in Ref. \cite{jeon2016protein}, we have computed the cumulative distribution $\Pi(r^{2},\Delta)$ for the squared displacements of the protein varying the time lag $\Delta$ as shown in Fig \ref{fig:cumulative_distr}. 
Here we recall that the cumulative distribution of the square displacements for the two dimensional motion is $\Pi(r^{2},\Delta)=\int_0^rP({\bf r'},\Delta)2\pi r'dr'$, where $P({\bf r},\Delta)$ is the propagator \cite{Weigel6438,schutz1997single,weiss2013single}. In the case of a Gaussian (anomalous) diffusion it takes the form $P({\bf r},\Delta)=\exp[-r^2/(2 \sigma_\Delta)]/(2 \pi \sigma_\Delta)$, with $\sigma_\Delta=2D_\alpha\Delta^\alpha$, which yields the cumulative distribution $\Pi(r^{2},\Delta)=1-\exp[-r^2/(4 D_\alpha\Delta^\alpha)]$. The plot of $-\log[1-\Pi(r^{2},\Delta)]$ versus $r^2$ in Fig.~\ref{fig:cumulative_distr} displays a reasonable power-law scaling with exponent 2, which supports the Gaussianity of the protein movement.

%%%%%%%%%%%%% Figure %%%%%%%%%%%%% 
%%%%%%%%%%%%% End Figure %%%%%%%%%%%%% 

The model has then been tested against the lateral MSD data extracted from the MD simulations. 
Two different flavors of the model have been implemented, corresponding to different choices of the free parameters $\boldsymbol{\theta}$:
\begin{enumerate}
\item[(M1)]  $\boldsymbol{\theta}= (\omega_s, \omega_p, \tau, \lambda, \delta)$, with constraint $\nu=\delta\lambda$;
\item[(M2)] $\boldsymbol{\theta}= (\omega_s, \omega_p, \tau,\lambda)$, with constraint $\delta=1$. Together with the condition $\delta\lambda=\nu$ (this leads to set $\nu=\lambda$). 
\end{enumerate}
Model (M1) corresponds to a three parameter Mittag-Leffler function, it is the more generic model, whilst model (M2) reduces to a two parameter Mittag-Leffler function, which has been already used in the literature in the context of viscoelasticity \cite{giusti2018prabhakar}.
The mass of the protein and the temperature are fixed to $M=41697$ g/mol and $T=310$ K in both implementations. The expected values of the parameters $\boldsymbol{\theta}$ and their uncertainties have been evaluated by using a Bayesian approach with a flat (uninformative) prior distribution for the parameters \cite{jaynes2003probability}, i.e. all the parameters values are assumed to be equally probable before analyzing the data.
Details on the data analysis are presented in Appendix~\ref{appex:statistics}.

%%%%%%%%%%%%% Figure %%%%%%%%%%%%% 
\begin{figure}[t]
  \centering
    \includegraphics[width=0.6\linewidth]{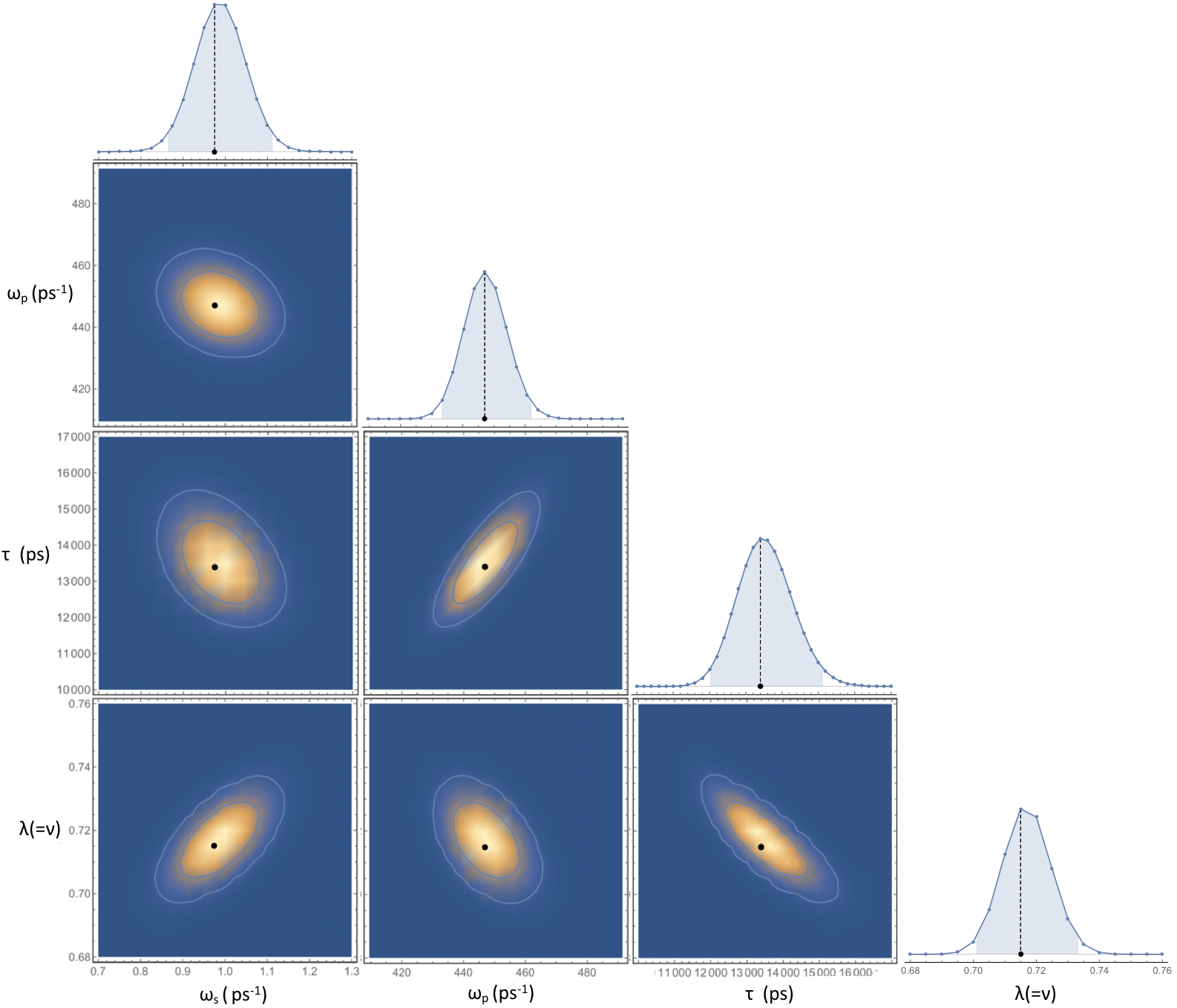}
%   \caption{One-dimensional and two-dimensional marginal distributions for model (M1) (Left) and model (M2) (Right). The highlighted area below the one-dimensional marginal distributions corresponds to 95\%. The ellipses in the 2-dimensional marginal distributions correspond to 68\% and 95\% confidence regions. The reduced $\tilde{\chi}^2$ is  0.94 and 1.02 respectively.}
\caption{One-dimensional and two-dimensional marginal distributions for model (M2). The highlighted area below the one-dimensional marginal distributions corresponds to 95\%. The ellipses in the 2-dimensional marginal distributions correspond to 68\% and 95\% confidence regions. The reduced $\tilde{\chi}^2$ is  0.94.}
    \label{1D2D}
\end{figure}
%%%%%%%%%%%%% End Figure %%%%%%%%%%%%% 

\begin{figure}[t]
    \centering
        \includegraphics[width=0.6\linewidth]{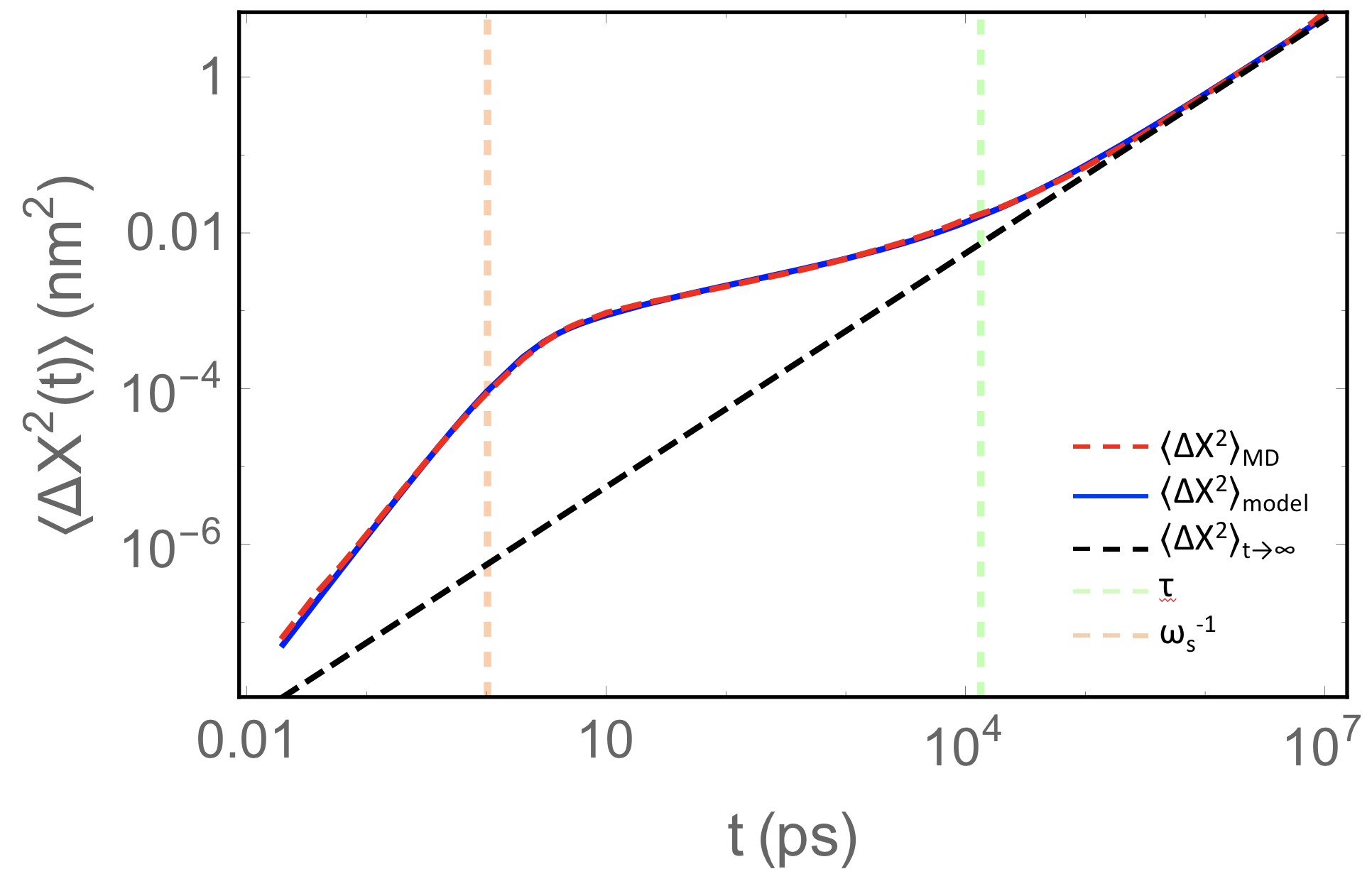}
        \caption{Comparison between the MSD data from MD simulations (red) and GLE model (blue). The long-time asymptotic MSD (black), and the characteristic times $\tau$ (green) and $\omega_{s}^{-1}$ (orange) are shown additionally.
        }
        \label{fig:comparison_MSD_ECM_MSD_MD_full_description}
\end{figure}

\begin{figure}[htp!]
\centering
        \includegraphics[width=0.6\linewidth]{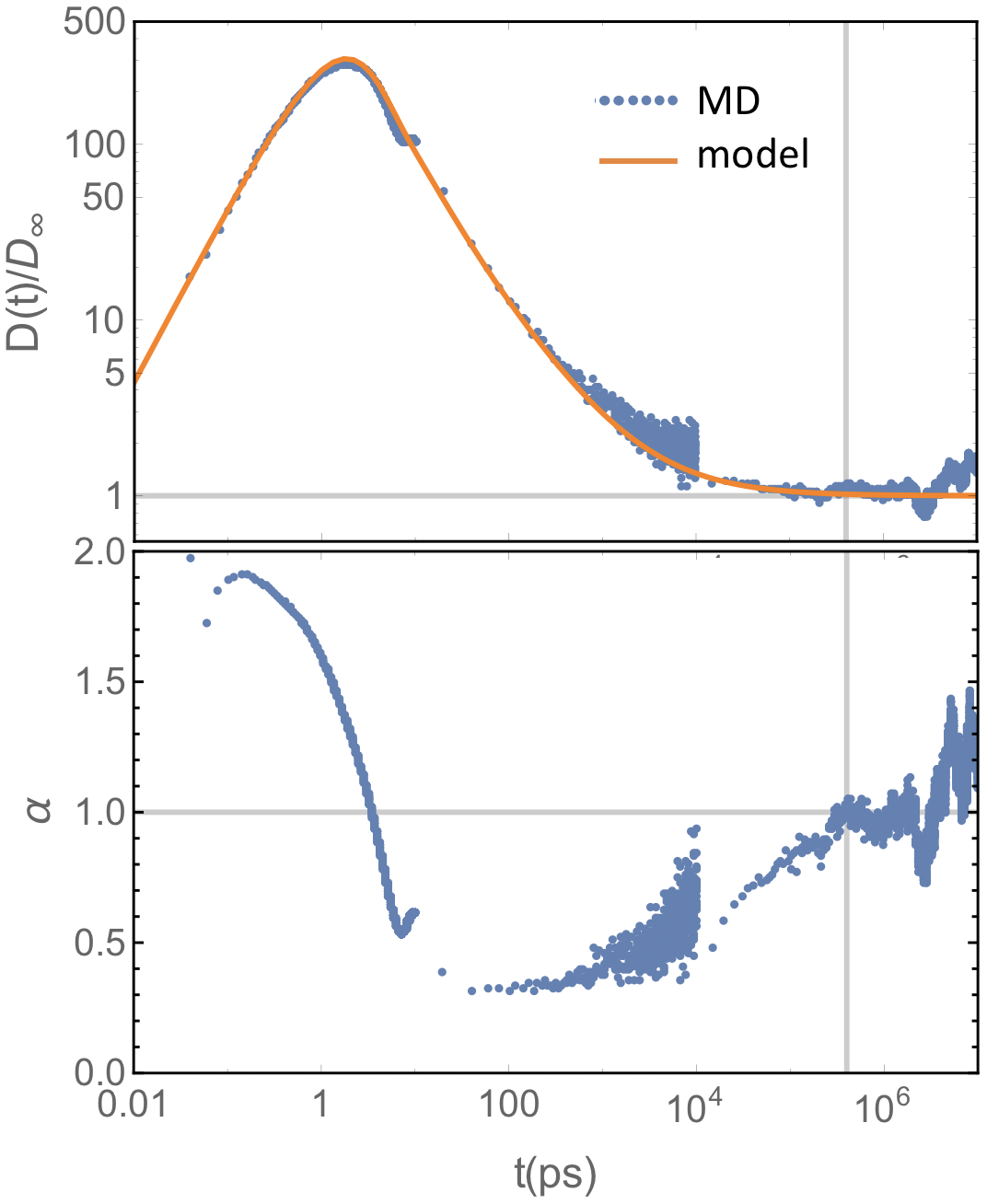}
        \caption{ Top: comparison between theoretical time-dependent diffusion coefficient normalized to the value $D_{\infty}=K_{B}T/(\xi_{s}+\xi_{p})$ and the numerical diffusion coefficient. Bottom: local MSD dependence on time. $\alpha=2$ indicates the ballistic regime, $\alpha=1$ the Brownian regime, $0<\alpha<1$ the subdiffusive regime. Gray lines help to visualize the onset of the Brownian dynamics.
        }
        \label{fig:DIFFCOEF_model_D(infinity)}
\end{figure}

\begin{figure}[htp!]
\centering
        \includegraphics[width=0.6\linewidth]{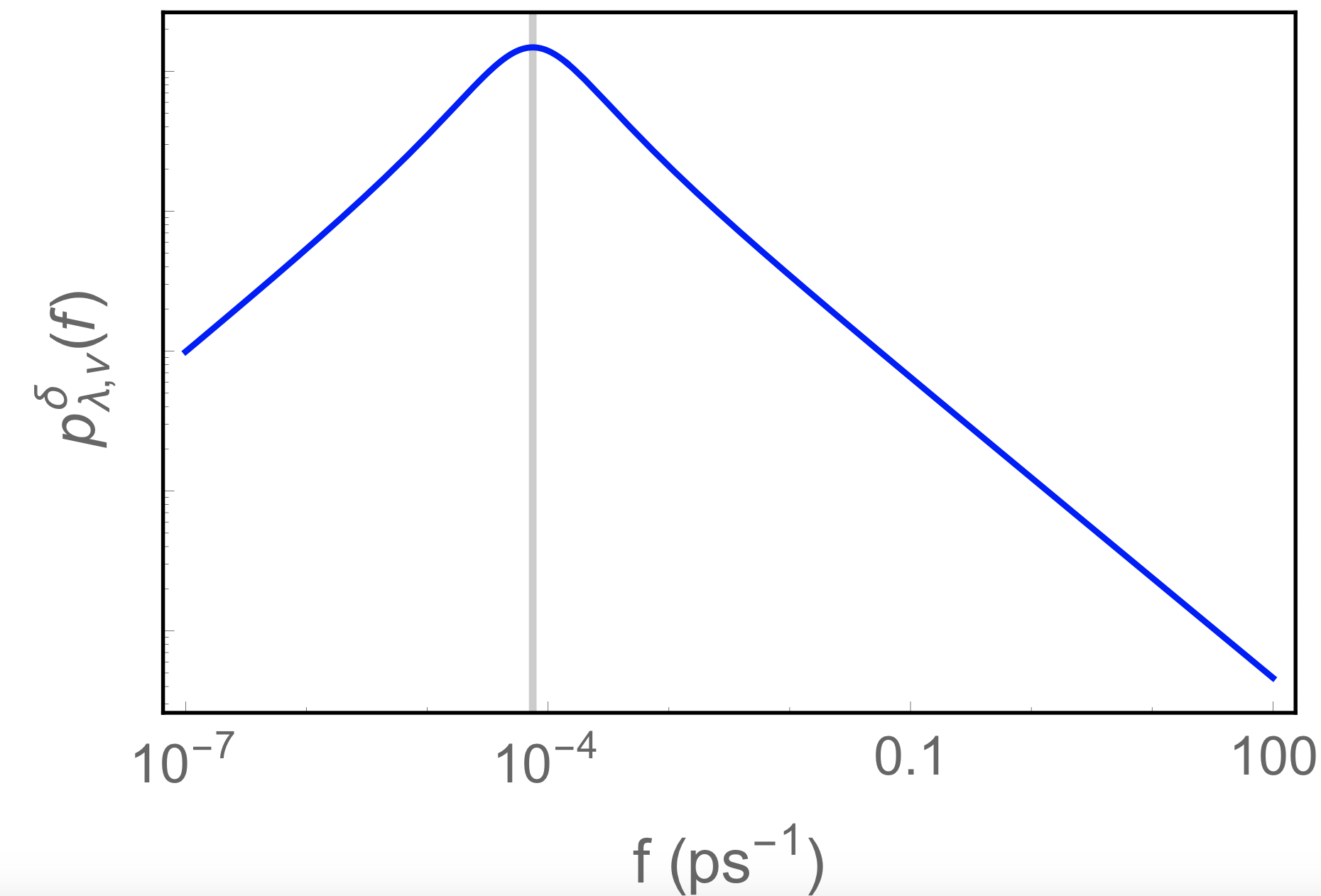}
        \caption{Distribution $p_{\lambda\nu}^{\delta}(f)$ of relaxation rates $f$ underlying the memory kernel. The gray line corresponds to $1/\tau$.}
        \label{fig:RelaxationRates}
\end{figure}

Due to the complexity of the MSD model in the time domain, Eq.~\eqref{def:stat_obs}, the model for a given set of parameters is first calculated in the Laplace space. Combining Eqs. \eqref{def:stat_obs}, \eqref{H_RF} and \eqref{zetaS}, under the condition $\delta\lambda=\nu$, it reads
\begin{equation}\label{eqn:modelfit}
    \langle \widehat{\Delta\bm{X}}^{2}\rangle(s)=4\frac{K_{B}T}{M}\frac{1}{s^2}\frac{1}{s+\omega_s+\omega_p\frac{1}{(1+(\tau s)^\lambda)^\delta}}.
\end{equation}
Then, in order to carry out the Bayesian analysis for the best fit parameters estimation, the inverse Laplace transform is applied to the latter using the numerical routine by Horváth et al. \cite{horvath2020numerical,inverseLT_package,inverseLT_link} in Mathematica for each set of sampled parameters.
\\
 Model (M1) was investigated by sampling the parameters space over a quite wide grid of values. This analysis shows that the best fit is obtained for $\delta\sim 1$, which points towards model (M2). We thus investigated model (M2) with a finer sampling in the parameter space (see Appendix~\ref{appex:statistics} for the grid of the investigated parameters values). Both models provide a reasonable fit in terms of reduced $\tilde{\chi}^2$, without significant changes for the best fit parameters. We thus report here in the main text the data and the analysis relevant to model (M2), while the data for model (M1) are available in Appendix~\ref{appex:statistics}. 
The parameters and the 95\% confidence intervals are summarized in Table \ref{tab:values_parameters} together with the parameters $\xi_s$ and $\xi_p$ derived thereof and the Brownian diffusion coefficient $D_\infty$. 
The one-dimensional marginalized posterior distributions, used to estimate the best fit values and the confidence intervals, are illustrated in Fig.\ref{1D2D} together with the two-dimensional marginalized posterior distributions, showing the correlations between the parameters. See Appendix~\ref{appex:statistics} for more details about their calculations.
Here we emphasize that the reported numerical values of the diffusion coefficient $D_\infty$ may suffer from finite size effects due to the periodic boundary conditions and the limited box size used in simulations \cite{Brown2015,Venable2017}. According with the error estimation reported in the literature for systems analogous to the one discussed here (i.e. protein radius $\sim 1.8$ nm, total water thickness 6.2 nm, lateral membrane edge 16.6 nm, Martini forcefield) an underestimation of $\sim 40\% $ is expected \cite{Brown2015,Venable2017}.
\\
The comparison in log-log scale between the model and the numerical MSD data is shown in Fig. \eqref{fig:comparison_MSD_ECM_MSD_MD_full_description}.
In order to highlight the physical role of the model parameters, we report in the same figure the asymptotic MSD obtained from the model $\langle\Delta \bm{r}^{2}(t)\rangle_{t\to\infty}=4\,D_\infty\, t$, with $D_\infty$ given by Eq. \eqref{eq:exactDinfty}, the characteristic time scale $\tau$ of the retarded response, and the transition time scale $1/\omega_s$ from the ballistic to subdiffusive regime.
\\
The evolution of the time dependent diffusion coefficient normalized to $D_\infty$ as obtained from the theoretical model and MD simulations is shown in Fig. \ref{fig:DIFFCOEF_model_D(infinity)}. Theoretical values are calculated from the second equation in \eqref{def:stat_obs} by inverse Laplace transforming \cite{horvath2020numerical,inverseLT_package,inverseLT_link} Eq. \eqref{H_RF}. Notice that in the subdiffusive regime, the diffusion coefficient is higher than in the Brownian regime (Fig.\ref{fig:DIFFCOEF_model_D(infinity)}). The coefficient $\alpha(t)$, expressing the momentary dependence of the MSD on time (MSD$\propto t^\alpha(t)$), as calculated from the logarithmic derivative of the simulated MSD data, is reported in the same figure.
\\
Both Figs. \ref{fig:comparison_MSD_ECM_MSD_MD_full_description} and \ref{fig:DIFFCOEF_model_D(infinity)} show that the subdiffusive to Brownian dynamics transition lasts from tens to hundreds of nanoseconds and that the Brownian dynamics is fully recovered only at time lags of $\approx$ 400 ns. Within this time-frame, the protein moves about $\approx$ 0.5 nm, to be compared with the radius of an average lipid, which is about $\approx$ 0.4 nm. This suggests that the subdiffusive behavior arises from the local interactions between the protein and the first lipid shell. The latter induces a ``transient trapping'' effect leading to a large distribution of relaxation times modeled through the retarded response of the system. 
The corresponding distribution function $p_{\lambda\nu}^\delta (f)$ for the relaxation rates $f$ can be obtained through the spectral representation of $(t/\tau)^{\nu-1}E_{\lambda,\nu}^\delta[(t/\tau)^\lambda]$ by Mainardi \cite{mainardi2015complete}:

\begin{equation}
\label{eqn:spectrum}
% \begin{aligned}
\left(\frac{t}{\tau}\right)^{\nu-1}E_{\lambda,\nu}^\delta\left[-\left(\frac{t}{\tau}\right)^\lambda\right]=\int_0^\infty p_{\lambda\nu}^\delta (f)e^{-f t}df,
\end{equation}
% & \text{where}\\
where
\begin{equation}
\begin{split}
p_{\lambda \nu}^\delta (f)&=\frac{\tau  (f  \tau )^{\lambda  \delta -\nu } \sin \left[\pi  (\nu -\lambda  \delta )+\delta \theta_\lambda(f)\right] }{\pi \left[(f \tau) ^{2 \lambda }+2 (f \tau) ^{\lambda } \cos (\pi  \lambda)+1\right]^{\delta/2}},
\\
\theta_\lambda(f)&=\arg
   \left[(-f \tau )^{\lambda }+1\right]
\end{split}
\end{equation}
for $0<\lambda\leq 1$ and $0<\lambda \delta\leq \nu\leq 1$.
%  \end{aligned}  
% \end{equation}
The calculated distribution for the optimally fitted parameters is shown in Fig.~\ref{fig:RelaxationRates}.

%%%%%%%%%%%%% Table %%%%%%%%%%%%% 
\begin{table*}[t]
\centering
\begin{tabular}{|c|c|c|c|c|c|c|c|}
\hline
$^\mathsection \delta$ & 
$^*\lambda$=$\nu$ & 
$^*\tau$ (ns) & 
$^*\omega_s$ (ps$^{-1})$ & 
$^*\omega_p$ (ps$^{-1})$ 
&
$^\dagger\xi_{s}$ (Pa $\cdot$ s$\,\cdot$ $\mu$m) &
$^\dagger\xi_{p}$ (Pa $\cdot$ s $\cdot \mu$m) & 
$^\dagger D_{\infty}$ ($\mu$m$^{2}\cdot$ s$^{-1}$) 
\\ \hline
$1$ & 
$0.715^{^{0.73}}_{_{0.70}}$ &
$13.4^{^{15.1}}_{_{12.0}}$ & 
$0.98^{^{1.11}}_{_{0.86}}$ &
$447^{^{462}}_{_{433}}$ &
$0.067^{^{0.077}}_{_{0.059}}\cdot 10^{-3}$ &
$ 0.0309^{^{0.0319}}_{_{0.0299}}$ &
$0.137^{^{0.142}}_{_{0.133}}$ 
\\
\hline
\end{tabular}
\caption{Best fit parameters ($^*$), fixed parameters ($^\mathsection$), and parameters derived from the best fit parameters ($^\dagger$) for model (M2). The sub- and superscripts represent the 95\% confidence intervals.}
\label{tab:values_parameters}
\end{table*}
%%%%%%%%%%%%% End Table %%%%%%%%%%%%% 

%%%%%%%%%%%%% Figure %%%%%%%%%%%%% 
%%%%%%%%%%%%% End Figure %%%%%%%%%%%%% 

%%%%%%%%%%%%% Figure %%%%%%%%%%%%% 
%%%%%%%%%%%%% End Figure %%%%%%%%%%%%% 

%%%%%%%%%%%%% Figure %%%%%%%%%%%%% 
%%%%%%%%%%%%% End Figure %%%%%%%%%%%%% 

\section*{Conclusion}
\label{sec:conclusion}
% {\color{green} The section Conclusion should appear separately as we did in the previous version}
%\section*{Conclusion}
%{\color{blue} BS:  there is conclusion and %discussion}

The proposed Generalized Langevin equation-based model, built on a memory kernel made of an instantaneous viscous component $\delta(t)$, and a retarded (elastic) component $t^{\delta\lambda-1}E_{\lambda,\delta\lambda}^{\delta}(-(t/\tau)^\lambda)$, proved to be able to describe the transition from the ballistic to the subdiffusive regime and from the subdiffusive to the Brownian regime, when the constraint $\nu=\lambda\delta$ is imposed. 
Notice that the model has been tested and validated at infinite protein dilution in a mixed membrane, while in crowded conditions deviations from the Gaussianity hypothesis, underlying the proposed model, have been reported \cite{jeon2016protein}. Our findings suggest that the transient subdiffusive behavior arises from the local interactions between the protein and the first lipid shell, inducing a ``transient trapping'' effect with a large distribution of relaxation times. A systematic study on the effect of different membrane compositions on this distribution of relaxation times is ongoing. The three-parameter Mittag-Leffler function has been already used in different contexts to model viscoelastic effects
\cite{PhysRevLett.102.058101,goychuk2012viscoelastic,de2011models}. In a continuum perspective, one could use this function to model the time-dependent (viscoelastic) membrane response within the constitutive  equation.
% {\green \sout{, and  thus derive a time-dependent lateral diffusion coefficient in terms of environment's properties such as the medium viscosity and the protein and membrane geometry}}. 

\section*{Acknowledgments}
The authors would like to offer their special thanks to Trifce Sandev for his advises about the theoretical analysis of our stochastic equation. The authors also would like to thank Riccardo Capelli and Luca Pesce for fruitful discussions and help in MD simulations setting.\\
This work was supported by the J\"ulich-Aachen Research Alliance Center for Simulation and Data Science (JARA-CSD) School for Simulation and Data Science (SSD). 
% \\
% {\color{green}I tentatively wrote the contributions, which are mandatory}
% \\
\section*{Author Contributions}
V.C. designed the project, L.D.C. and B.S. performed analytical derivation, L.D.C. performed MD simulations, L.D.C. and V.C. performed the analysis, L.D.C, B.S. and V.C. wrote the manuscript. 

\section*{Declaration of Interests}

The authors declare no competing interests.

\appendix

\section{Three parameter Mittag-Leffler function}
\label{appex:ML}

\begin{figure}[htp!]
\centering
        \includegraphics[width=0.6\linewidth]{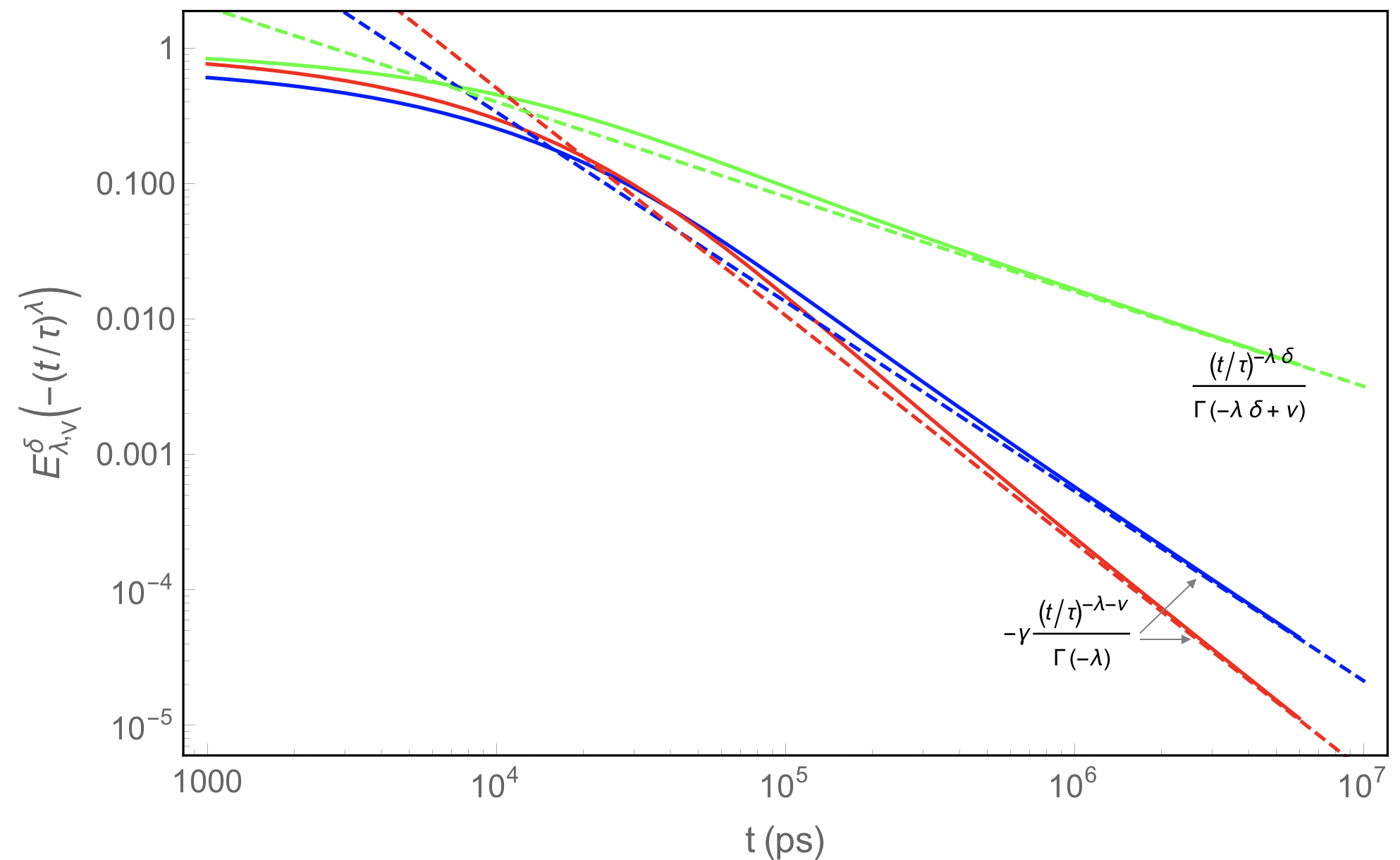}
        \caption{Illustration of the three parameter Mittag-Leffler function (solid lines) for three different set of parameters (blue: as presented in Table~\ref{fig:comparison_MSD_ECM_MSD_MD_full_description}; red: $\tau=13.4 ns$, $\lambda=0.7$, $\nu=0.98$, $\delta=1.4$; green: $\tau=13.4 ns$, $\lambda=0.7$, $\nu=1.0$, $\delta=1.0$) as well as their corresponding long time asymptotic decay (dashed lines). The numerical representation is based on the spectral distribution of the Prabhakar function reported in Ref. \cite{mainardi2015complete}.} 
    \label{fig:long_3ML}
\end{figure}

The three-parameter Mittag-Leffler function is defined as \cite{prabhakar1971singular,convergence_mittag_leffler}
\begin{equation}\label{def_ML_3par}
    E^{\delta}_{\lambda,\nu}(z)=\sum_{k=0}^{\infty}\frac{(\delta)_{k}}{\Gamma(\lambda k+\nu)}\frac{z^{k}}{k!},
\end{equation}
where $(\delta)_{k}$ the Pochhammer symbol whereas $\Gamma$ is the Euler-gamma function and $\lambda, \,\nu,\,\delta$ could be, in general, complex numbers but with $Re[\lambda]>0$.
\\
The long time expression of the three-parameter Mittag-Leffler function follows a power-law behavior given by \cite{ift_vacf_hh}:
\begin{equation}\label{Eq:asymptotic_ML_3para}
    E^{\delta}_{\lambda,\nu}(-z)\approx \frac{z^{-\delta}}{\Gamma(\nu-\delta\lambda)}-\frac{z^{-(\delta+1)}}{\Gamma(\nu-(\delta+1)\lambda)}, \qquad (|z|\gg 1).
\end{equation}
A few different examples of Mittag-Leffler functions are illustrated in Figure \ref{fig:long_3ML} together with their asymptotic behavior.  
The plots are given for a set of parameters verifying the complete monotonicity condition, $0<\lambda\leq 1$, $0<\lambda\delta\leq\nu\leq 1$, and are derived from the spectral distribution of $t^{\nu-1}E^{\delta}_{\lambda,\nu}(-(t/\tau)^{\lambda})$ by Mainardi \cite{mainardi2015complete}.

%%%%%%%%%%%%% Figure %%%%%%%%%%%%% 
%%%%%%%%%%%%% End Figure %%%%%%%%%%%%% 

\section{Numerical details}
\label{appex:numerics}
In this appendix, we report further details of the MD simulation setup. 
The details about the constituents of the system, the membrane composition, the geometrical parameters of the system, and the trajectories length are reported in Tables \ref{tab:numb_water_ions}, \ref{tab:numb_lipids}, \ref{tab:prot_dimensions}, and \ref{tab:sampling_sim}, respectively.

%%%%%%%%%%%%% Table %%%%%%%%%%%%% 
\begin{table}[t]
\centering
\begin{tabular}{|l|c|c|}
\hline
\textbf{Lipid Type} & \textbf{Upperleaflet Nº}& \textbf{Lowerleaflet Nº} \\
\hline
\text{CHOL} & 300 & 300\\
\text{DOPC} & 12 & 12\\
\text{DPPC} & 30 & 30\\
\text{POPC} & 54 & 54\\
\text{DGPE} & 12 & 12\\
\text{POPS} & 18 & 18\\
\text{PAPI}& 18 & 18\\
\text{DPSM} & 54 & 54\\
\text{PNSM} & 18 & 18\\
\text{POSM} & 6 & 6\\
\hline
\end{tabular}
\caption{The number of lipids for each species composing the membrane.}
\label{tab:numb_lipids}
\end{table}
%%%%%%%%%%%%% End Table %%%%%%%%%%%%% 

%%%%%%%%%%%%% Table %%%%%%%%%%%%% 
\begin{table}[t]
\centering
\begin{tabular}{|l|c|c|c|c|}
\hline
\text{Nº of water molecules} & $14077$\\
\hline
\text{Nº of Na$^+$ ions} & $825$ \\
\hline
\text{Nº of Cl$^-$ ions} & $161$\\
\hline
\text{Nº of lipids}  & $1044$ \\
\hline
\text{Nº of proteins}  & $1$ \\
\hline
\end{tabular}
\caption{The number of each constituent of the system.}
\label{tab:numb_water_ions}
\end{table}
%%%%%%%%%%%%% End Table %%%%%%%%%%%%% 

%%%%%%%%%%%%% Table %%%%%%%%%%%%% 
\begin{table}[t]
\centering
\begin{tabular}{|l|c|c|}
\hline
\textbf{} & \textbf{Upperleaflet}& \textbf{Lowerleaflet} \\
\hline
\text{Protein Area} & 1152.12249 $\mbox{Å}^{2}$ & 1126.52528 $\mbox{Å}^{2}$\\ 
\hline
\text{Lipid Area} & 26331.6 $\mbox{Å}^{2}$ & 26331.6 $\mbox{Å}^{2}$\\
\hline
\text{Total Area} & 27483.72249 $\mbox{Å}^{2}$ & 27458.12528 $\mbox{Å}^{2}$\\
\hline 
\hline
\text{Protein X Extent} & 26.06 $\mbox{Å}$ &\\
\hline
\text{Protein Y Extent} & 24.69 $\mbox{Å}$ &\\
\hline
\text{Box Y Extent} & 165.74 $\mbox{Å}$ &\\
\hline
\text{Box X Extent} & 165.74 $\mbox{Å}$ & \\
\hline
\text{Box Z Extent} & 109.31 $\mbox{Å}$ & \\
\hline
\end{tabular}
\caption{Geometrical parameters of the simulated system.}
\label{tab:prot_dimensions}
\end{table}
%%%%%%%%%%%%% End Table %%%%%%%%%%%%% 

%%%%%%%%%%%%% Table %%%%%%%%%%%%% 
\begin{table}[t]
\centering
\begin{tabular}{|l|c|c|}
\hline
\text{Regimes} & \text{Sampling time step} & \text{Time length}\\
 \hline
 \text{Ballistic} & $20$ fs & $1000$ ps\\
 \hline
 \text{Subdiffusive} & $20$ ps &  $3\,\mu$ s\\
 \hline
 \text{Diffusive} & $5000$ ps & $100\,\mu$s\\
\hline
\end{tabular}
\caption{Sampling time step and total length of the numerical simulations.}
\label{tab:sampling_sim}
\end{table}
%%%%%%%%%%%%% End Table %%%%%%%%%%%%% 

The whole system was prepared by using CHARMM-GUI membrane builder web server \cite{bworld}. The simulations were performed through the GROMACS program suite \cite{berendsen1995gromacs,lindahl2001gromacs,van2005gromacs,pronk2013gromacs}, version 2016.3, in double precision, using a Martini force field \cite{qi2015charmm,marrink2007martini,marrink2004coarse} with the leap-frog integration algorithm \cite{pall2013flexible} and an integration time step of $20$ fs. Electrostatic interactions were calculated using the reaction-field scheme and a cutoff of $1.1$ nm for both the electrostatic and van der Waals interactions was used.\\
We followed the minimization and equilibration protocol of CHARMM-GUI previously used in our lab on analogous systems \cite{Capelli2021}:
\begin{enumerate}
    \item Minimization
    \begin{itemize}
        \item $2.5\cdot 10^{5}$ steps of Steepest descent minimization with a soft-core potential on LJ and Coulomb interaction;
        \item $2.5\cdot 10^{4}$ steps of Steepest descent minimization;
    \end{itemize}
    \item The equilibration is carried out in NPT ensemble using the Berendsen barostat \cite{di1994molecular} (reference pressure 1 bar, coupling constant 5 ps, and  compressibility $3 \cdot 10^{-4}\; \mbox{bar}^{-1}$) and the velocity rescale thermostat \cite{bussi2007canonical} (reference temperature 310 K, coupling constant 1 ps). The procedure is splitted in 5 simulation steps:
    \begin{itemize}
    \item $5\cdot 10^{5}$ steps (time step $0.002$ ps, trajectory length 1000 ps) with position restraint on the head groups (200 kJ/mol/nm$^{2}$) and the protein (1000 kJ/mol/nm$^{2}$);
    \item $2\cdot 10^{5}$ steps (time step $0.005$ ps, trajectory length 1000 ps) with position restraint on the head groups (100 kJ/mol/nm$^{2}$) and the protein (500 kJ/mol/nm$^{2}$);
    \item $1\cdot 10^{6}$ steps (time step 0.010 ps, trajectory length 10000 ps) with position restraint on the head groups (50 kJ/mol/nm$^{2}$) and the protein (250 kJ/mol/nm$^{2}$);
    \item $5\cdot 10^{5}$ steps (time step 0.015 ps, trajectory length 7500 ps) with  position restraint on the head groups (20 kJ/mol/nm$^{2}$) and the protein (100 kJ/mol/nm$^{2}$);
    \item $5\cdot 10^{5}$ steps (time step 0.020 ps, trajectory length 7500 ps) with position restraint on the head groups (10 kJ/mol/nm$^{2}$) and the protein (50 kJ/mol/nm$^{2}$);
    \end{itemize}
\end{enumerate}
Three production runs were carried out in NPT ensemble using the Parrinello-Rahman barostat \cite{parrinello1981polymorphic} (reference pressure 1 bar, coupling constant 15 ps, compressibility $3 \cdot 10^{-4} \;\mbox{bar}^{-1}$) and the velocity rescale thermostat \cite{bussi2007canonical} (reference temperature 310 K, coupling constant 1 ps). Different sampling times were used in order to fully resolve the different dynamical regimes (See Table~\ref{tab:sampling_sim}).
From these simulations we extracted the protein's center-of-mass trajectories.
The lateral MSD of the center of mass was calculated through the Gromacs tool {\tt analyze} with the flag {\tt-msd}. The maximum time lag considered corresponds to $10 \%$ of the trajectory length.
\\
To give an idea of the computational effort, a 10 µs-long trajectory carried out on a 48-cores machine (Intel Xeon(R) CPU E5-2687W v4) requires 4 days (i.e. 2.5 µs/day).

\section{Data Analysis}
\label{appex:statistics}

In order to evaluate best fit values of the model parameters and their uncertainties, we used a Bayesian approach with a flat (uninformative) prior distribution for the parameters \cite{jaynes2003probability}, i.e. all the parameters values are assumed to be equally probable before analyzing the data.
\\
According with the Bayes's theorem, the \emph{posterior} distribution of a model described by the parameters $\boldsymbol{\theta}$, given the measured data $\mathbf{d}$, reads:
\begin{equation}
\mathcal{P}(\boldsymbol{\theta}|\mathbf{d}) = \frac{\mathcal{L}(\mathbf{d}|\boldsymbol{\theta})\Pi(\boldsymbol{\theta})}{E(\mathbf{d})},
\end{equation}
where
\begin{equation}
\mathcal{L}(\mathbf{d}|\boldsymbol{\theta}) = \frac{1}{Z}\mathrm{e}^{-\chi^2/2} = \frac{1}{Z} \exp \left[-\frac{1}{2}\sum_{i=1}^N\left(\frac{m(\boldsymbol{\theta},i)-d_i}{\sigma_i}\right)^2\right],
\end{equation}
\\
is the \emph{likelihood} of the MSD data $\boldsymbol{d}=\{d_i\}\,$, given the  MSD model  $m(\boldsymbol{\theta},i)\,$ calculated for the parameters $\boldsymbol{\theta}$, with $i$ being the index running over the points of the data set of size $N\,$, and $\sigma_i$ the error of the data. $Z$ is a normalization constant that guarantees that $\int\mathcal{L}(\mathbf{d}|\boldsymbol{\theta})d^5\boldsymbol{\theta}=1$.  The \emph{evidence} of the data, $E(\mathbf{d})$, amounts to a normalization constant whose value is fixed and $\Pi(\boldsymbol{\theta})$ is the \emph{prior} distribution of parameters, which is assumed to be uniform \cite{jaynes2003probability}.\\
To obtain an accurate estimation of the subdiffusive to Brownian dynamics crossover, many microscopic fluctuations must be observed. Thus, the relative error on the observed MSD has been estimated as the reciprocal of the square root of the number of observed fluctuations and so as $(\tau^*/T)^{1/2}$, where $\tau^*$ is the characteristic time of the fully Brownian dynamics recovering, $\approx 400$ ns, and $T$ is the observation time (i.e. the largest trajectory length).\\
The posterior distribution of the model is sampled over a regular grid of the parameter space:

\begin{enumerate}
\item[M1:]  $\omega_s \in [\,0.6,1.4]\,$ ps$^{-1}$, $\omega_p \in [\,408,500]\,$ ps$^{-1}$, $\tau\in [\,7000,20000]\,$ ps, $\lambda \in [\, 0.4,1.0] \,$, and $\delta \in [\, 0.5,1.5]\,$;
\item[M2:] $\omega_s \in [\,0.6,1.2]\,$ ps$^{-1}$, $\omega_p \in [\,382.2,461.1]\,$ ps$^{-1}$, $\tau \in [\,10900,16900] \,$ ps, and $\lambda(=\nu) \in [\, 0.66,0.74] \,$. 
\end{enumerate}

\begin{figure}[htp!]
  \centering
    \includegraphics[width=8.5cm,scale=1.5]{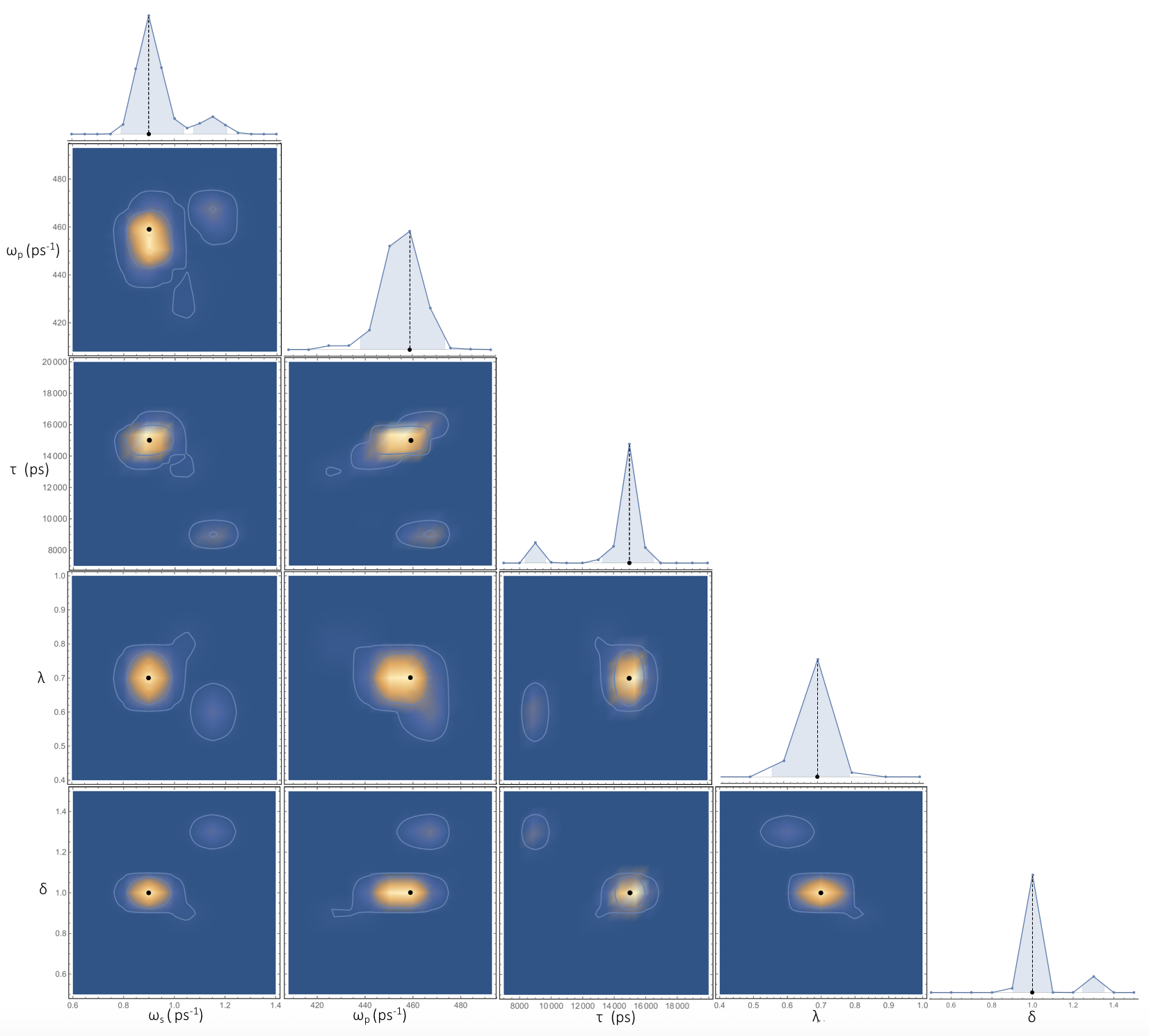}
  \caption{One-dimensional and two-dimensional marginal distributions for model (M1). The highlighted area below the one-dimensional marginal distributions corresponds to 95\%. The ellipses in the 2-dimensional marginal distributions correspond to 68\% and 95\% confidence regions. The reduced $\tilde{\chi}^2$ is 1.02.}
    \label{1D2D5param}
\end{figure}

The expectation value of each parameter is estimated by the maximum of the one-dimensional marginalized posterior distribution, i.e. after marginalizing $\mathcal{P}$ over all but one parameter in turn (Fig. \ref{1D2D} for model (M2) and Fig.\ref{1D2D5param} for model (M1)).
The parameters confidence intervals correspond to the 95\% of the area below the one-dimensional marginalized posterior distribution starting from the maximum. The best fit parameters are summarized in Table\ref{tab:values_parameters} for model (M2) and Table \ref{tab:values_parameters2} for model (M1).
In order to estimate the correlations between the parameters of the model, the $2$-dimensional marginal distribution for each pair of parameters (obtained by marginalizing over $n-2$ parameters, where $n$ is the number of free parameters) is calculated as well (Fig. \ref{1D2D} for model (M2) and Fig.~\ref{1D2D5param} for model (M1)).

%%%%%%%%%%%%% Table %%%%%%%%%%%%% 
\begin{table*}[t]
\centering
\begin{tabular}{|c|c|c|c|c|c|c|c|}
\hline
$^*\delta$ & 
$^*\lambda$=$\delta\nu$ & 
$^*\tau$ (ns) & 
$^*\omega_s$ (ps$^{-1})$ & 
$^*\omega_p$ (ps$^{-1})$ 
\\ \hline
$1.0^{^{1.3}}_{_{0.9}}$ & 
$0.7^{^{0.8}}_{_{0.55}}$ &
$15.0^{^{16.5}}_{_{8.5}}$ & 
$0.9^{^{1.2}}_{_{0.8}}$ &
$459^{^{475}}_{_{440}}$ 
\\
\hline
\end{tabular}
\caption{Best fit parameters ($^*$) of model (M1)
The sub- and superscripts represent the 95\% confidence intervals.}
\label{tab:values_parameters2}
\end{table*}
%%%%%%%%%%%%% End Table %%%%%%%%%%%%% 

%%%%%%%%%%%%% Figure %%%%%%%%%%%%% 
%%%%%%%%%%%% End Figure %%%%%%%%%%%%% 

% Uncomment if using bibtex (default)
\bibliography{aipsamp}

% Uncomment if using biblatex
% \printbibliography

\end{document}